\documentclass{article}
\usepackage[round]{natbib}
\usepackage{natbib}
\usepackage{amssymb}
\usepackage{epsfig}
\newlength{\FIGSIZE}
\newlength{\figsize}
\usepackage{color}

\newcommand{\be}{\begin{equation}}
\newcommand{\ee}{\end{equation}}
\newcommand{\ba}{\begin{eqnarray}}
\newcommand{\ea}{\end{eqnarray}}

\newcommand{\fett}{}

\newcommand{\lr}[1]{\langle #1 \rangle}

\newcommand{\bi}[1]{Fig.~\ref{fig:#1}}
\newcommand{\bis}[1]{Figs.~\ref{fig:#1}}
\newcommand{\e}[1]{Eq.~(\ref{#1})}
\newcommand{\es}[1]{Eqs.~(\ref{#1})}

\newcommand{\bd}{\bar{D}}
\newcommand {\var}{$\sigma^2$}
\newcommand {\la}{\langle}
\newcommand {\ra}{\rangle}

\begin{document}

\title{Are the input parameters of white-noise-driven integrate \& fire neurons uniquely determined by rate and CV?}
\author{Rafael D. Vilela and Benjamin Lindner\\
Max-Planck-Institut f\"ur Physik Komplexer Systeme\\
N\"othnitzer Str.~38, 01187 Dresden, Germany}
\date{\today}

\maketitle


\setlength{\baselineskip}{1em}

\begin{abstract}
Integrate \& fire (IF) neurons have found widespread applications in
computational neuroscience. Particularly important are stochastic versions of
these models where the driving consists of a synaptic input modeled as white
Gaussian noise with mean $\mu$ and noise intensity $D$. Different IF models
have been proposed, the firing statistics of which depends nontrivially on the
input parameters $\mu$ and $D$.
In order to compare these models among each other, one must first specify the
correspondence between their parameters.  This can be done by determining
which set of parameters ($\mu$, $D$) of each model is associated to a given
set of basic firing statistics as, for instance, the firing rate and the
coefficient of variation (CV) of the interspike interval (ISI).
However, it is not clear {\em a priori} whether for a given firing rate and CV
there is only one unique choice of input parameters for each model. Here we
review the dependence of rate and CV on input parameters for the perfect,
leaky, and quadratic IF neuron models and show analytically that indeed in
these three models the firing rate and the CV
uniquely determine the input parameters.   
\end{abstract}






\section{Introduction}\label{intr}
Stochastic integrate \& fire (IF) neurons constitute an important tool in
theoretical neuroscience, having been used to address a number of relevant
biological problems.  For instance, different variants of these models have
been employed in the debate on the high variability of the interspike interval
(ISI) observed for cortical neurons \citep{SofKoc93,GutErm98}. Other problems
in which stochastic IF models have been applied include the response to fast
signals \citep{BruCha01, LinLSG01,FouHan03,NauGei05}, asynchronous spiking in
recurrent networks \citep{Bru00}, and oscillations of firing activity in
systems with spatially correlated noisy driving \citep{DoiCha03, DoiLin04,
LinDoi05}.

IF neurons can be classified according to the nonlinearities that govern their
subthreshold dynamics. Three simple and important variants are the perfect
(PIF), leaky (LIF), and quadratic (QIF) models, in which the subthreshold
dynamics is described by a constant, a linear, and a quadratic voltage
dependence, respectively. Noisy inputs with different degrees of biological
realism have been considered for these models. One simple choice is a white
Gaussian input current, corresponding to the so-called diffusion approximation
of synaptic spike train input \citep{Hol76,Ric77,Tuc89}.  The PIF with white
noise drive (also referred to as the random walk model of neural firing) was
first considered by \cite{GerMan64}. The LIF with a white Gaussian input
current has been studied by \citep{Joh68} and afterwards by many other authors
(see \cite{Hol76,Ric77,Tuc89,Bur06} and references therein); it is also
referred to as Ornstein-Uhlenbeck neuron \citep{LanRos95}. White-noise driving
in the QIF was studied by \cite{GutErm98} and \cite{LinLon03}.


The firing statistics of the various IF models may depend sensitively on the
specific nonlinearity in the respective model , and so it is not clear at the
first glance which model is capable of reproducing which features of the
firing statistics of real neurons. For instance, it has been shown that LIF
and QIF display rather different phase shifts if driven by a periodic stimulus
\citep{FouHan03}. Further, the LIF with periodic stimulation can transmit
signals of very high frequencies if they are encoded in the noise intensity
\citep{LinLSG01}, while the QIF cannot \citep{NauGei05}. Last but not least, when the
strength of the input noise is varied, the LIF can show coherence resonance
(CR) \citep{PakTan01,LinLSG02} whereas the PIF and the QIF do not display CR
\citep{LinLon03} (see also below). A thorough comparison of different
stochastic IF neurons and the roles of their respective nonlinearities
constitutes therefore an interesting and largely open problem.

If one wants to compare two specific IF models one should tune their input
parameters (mean and intensity of the {\fett input} fluctuations) such that
the basic firing statistics is the same. A simple choice for the basic firing
statistics is the firing rate, quantifying the intensity of the spike train, and
the interspike interval's coefficient of variation (CV), quantifying the
irregularity of the spike train. Setting both models in this way in the same
firing regime (e.g. low spike rate and high ISI variability), one can then ask
for higher statistics (e.g. the power spectrum of the spike train) or for
their response characteristics (e.g. to weak periodic stimulation or step
currents). The idea for such a tuning tacitly assumes that there is for each
of the models at most one input parameter set yielding a desired firing
statistics, for instance, rate and CV. At a closer look, however, this is not
evident at all: are the input parameters for a white-noise driven IF model
indeed uniquely determined if we prescribe certain values of the rate and the
CV? This is the question that we address in this paper and we will answer it
for the three IF models mentioned above, namely, PIF, LIF, and QIF neurons.

The question discussed here is also related to the problem of parameter
estimation for an IF model from experimental data, which has been subject of
several studies (see \cite{LanDit08} and references therein). In one approach,
the model parameters are inferred from subthreshold membrane measurements (for
a recent reference see \cite{BadLef08}).  In another approach, model
parameters are estimated using solely the ISI statistics
\citep{TucRic78,InoSat95,RauLac03,LacRau04, ShiSak99, DitLan05, DitLan07,
MulIye08}. The latter approach is of importance, since often subthreshold data
are not available. Most of these studies consider the leaky IF model. To the
best of our knowledge, the problem of the uniqueness of input parameters has
been addressed only by \cite{KosLan07}, where it is mentioned without a
proof that uniqueness of the input parameters given fixed rate and CV holds
for the LIF.

In this paper, we show analytically that rate and CV uniquely determine the
input parameters for the three models. We first introduce the models and the
statistics studied here (sec.~\ref{sec:models}). In the main part of the paper
(secs.~\ref{sec:pif}-\ref{sec:qif}), we briefly review the rate and CV as
functions of the input parameters and then prove the uniqueness of the
relation between these parameters and the former statistics. We give an
outlook to generalizations of the considered problem in sec.~\ref{sec:conc}.


\section{Integrate \& fire neuron models}\label{models}
\label{sec:models}
\subsection{Definition of the models and relation to the first passage time problem}
IF models consist of two ingredients: (i) a one dimensional stochastic
ordinary differential equation describing the subthreshold behavior of the
membrane potential $V$ as a function of time $t$ and (ii) a fire-and-reset
rule.  The equation for the membrane potential has the form of a current-balance equation \citep{Bur06}:
\be 
C_m
\frac{dV}{dt}=I_{\mbox{\tiny{model}}}(V)+I_{\mbox{\tiny{syn}}}(t)+I_{\mbox{\tiny{ext}}}(t),
\label{if_orig}
\ee
where $C_m$ is the membrane capacitance, $I_{\mbox{\tiny{syn}}}(t)$ and
$I_{\mbox{\tiny{ext}}}(t)$ denote the synaptic and injected current,
respectively.  Here we will consider the case where
$I(t)=I_{\mbox{\tiny{syn}}}(t)+I_{\mbox{\tiny{ext}}}(t)$ is a white Gaussian
noise current with a constant mean value $\la I \ra$ and a correlation
function $\la (I(t)-\lr{I})(I(t')-\lr{I}) \ra = 2D_I \delta(t-t')$. The function $I_{model}(V)$
stands for a model-specific current --- for instance, a passive leak of the
membrane would be described by setting $I_{model}=-[V-V_0 ]/R_m$, where $V_0$
and $R_m$ denote the leak reversal potential and the passive membrane resistance
constant, respectively.  

The fire-and-reset rule can be expressed as
\be
V(t)=V_{th} \Longrightarrow \mbox{  spike at time $t$ and  } V \rightarrow V_r ,
\label{reset}
\ee
i.e., whenever the membrane potential reaches a threshold value $V_{th}$ the
neuron fires a spike and there is a reset of its membrane potential to a value
$V_r$.

It is convenient to make the following changes of variables:
\be
v=\frac{V-V_r }{V_{th}-V_r}, \hspace{4mm} t \to \frac{t}{\tau_m},
\label{change}
\ee
where $\tau_m =C_m R_m$ is the membrane time constant and the new variables,
$v$ and $t$, are dimensionless. This procedure corresponds to measuring the
voltage in units of the difference between threshold and reset (with $V_r$ as
the reference voltage) and time in units of the membrane time constant.

Defining
\be
f_{\mbox{\tiny{model}}}=\frac{R_m } {(V_{th}-V_r )}\left[I_{\mbox{\tiny{model}}} ((V_{th}-V_r )v
+V_r )-I_{\mbox{\tiny{model}}} (V_r)\right],
\ee
\be
\label{mu_I}
\mu=\frac{R_m}{(V_{th}-V_r )} \left[\la I \ra+I_{\mbox{\tiny{model}}} (V_r)\right],
\ee
and
\be
\label{D_I}
D=\frac{D_I R_{m}^{2} }{\tau_m (V_{th}-V_r )^2},
\ee
we can recast \e{if_orig} into the form:
\be
\dot{v}=f_{\mbox{\tiny{model}}}(v)+\mu+\sqrt{2D}\xi(t), 
\label{if} 
\ee
which is the equation that we will work with in the remainder of this
work\footnote{Note that, by construction,
$f_{\mbox{\tiny{model}}}(v_R)=0$. Another choice for $v$ is possible that also
leads to \e{if} in which $v$ denotes the deviation of the membrane potential
from the leak reversal potential $V_0$ instead of the scaled deviation of the
membrane potential from the reset potential $V_r$ as in our case.}. The
parameters $\mu$ and $D$ are called {\it input parameters}, i.e. they
represent the mean and the intensity of the fluctuating input in our
nondimensional model. Note that they are linearly related to the physiological
input parameters $\lr{I}$ and $D_I$ via \e{mu_I} and \e{D_I}. The function
$\xi(t)$ is a zero-mean white Gaussian noise with $\la \xi(t)
\xi(t') \ra=\delta(t-t')$. In the nondimensional formulation \e{if}, the reset and
threshold values, $v_r$ and $v_{th}$, are given by zero and one,
respectively. For better applicability of our results, we will keep $v_r$ and
$v_{th}$ in all resulting formulas and equations.

If one neglects any voltage dependence of the right-hand side in \e{if}, one
deals with a perfect integrator and the respective IF model is called a
perfect integrate \& fire (PIF) neuron: \be f_{\mbox{\tiny{PIF}}}=0,
\hspace{8mm} \mu > 0, \hspace{8mm} v_{th}=1,
\hspace{8mm} v_r=0. \label{f_pif} \ee 
If a leak current is taken into account, $f(v)$ becomes a linear function (by
construction, any additive constant is lumped into the mean input $\mu$). In
this case, we deal with a leaky integrate \& fire (LIF) neuron which is the
most often studied and used IF model to date:
\be
f_{\mbox{\tiny{LIF}}}=-v, \hspace{8mm} \mu \in (-\infty, \infty), 
\hspace{8mm} v_{th}=1, \hspace{8mm} v_r=0.
\label{f_lif}
\ee

Another IF model, the quadratic integrate \& fire (QIF) model also fits into
the framework of \e{if} although in its derivation $v$ is not a voltage but a
more abstract variable.  Higher nonlinearities in the current-balance
\e{if_orig} come along by voltage-dependent conductances and additional
variables like gating variables (e.g. in the classical Hodgkin-Huxley model)
or effective recovery variables (e.g. in the Morris-Lecar model). A wide class
of neurons (type I neurons) described by such multidimensional dynamical systems
shows a behavior associated with a saddle-node bifurcation \citep{RinErm89} in
these multidimensional models. For these neurons, the dynamics close to the
bifurcation is governed by one slow variable only which is the variable $v$ in
our general IF model \e{if}. The nonlinearity corresponding to the normal form
of a saddle-node bifurcation is a quadratic function. Furthermore, for this
specific dynamics the behavior is governed by the slow motion close to $v=0$
and threshold as well as reset values do not matter if taken far away from the
origin. For simplicity, they are commonly taken at infinity. Thus the dynamics
of a QIF model is described by the following setup:
\be
f_{\mbox{\tiny{QIF}}}=v^2, \hspace{8mm} \mu \in (-\infty, \infty). \hspace{8mm} v_{th}=\infty,
\hspace{8mm} v_r=-\infty.
\label{f_qif}
\ee
%

It is useful to interpret \e{if} as describing a Brownian particle of
position $v$ undergoing overdamped motion in a potential
$U{\mbox{\tiny{model}}}$ such that
\be
-\frac{dU{\mbox{\tiny{model}}}}{dv}=f_{\mbox{\tiny{model}}} + \mu.  
\label{potentials}
\ee
\begin{figure}[h!]
\centerline{\includegraphics[width=0.45\textwidth]{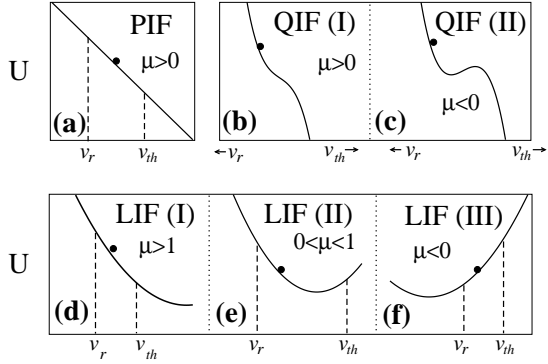}}
\caption{
\label{fig:pot}
Potentials for the different models (cf. \e{potentials}).  The PIF (a) is only
defined in the tonic firing regime ($\mu >0$) while the QIF ((b) and (c)) can be
in the tonic (QIF I in (b)) or in the noise-induced (QIF II in (c)) firing
regimes. The LIF displays tonic firing (LIF I in (d)) for $\mu>1$ and two
different noise-induced firing regimes ((e) and (f)) for which we distinguish the
cases where the potential minimum lays between $v_r$ and $v_{th}$ (e) and
where the minimum is to the left of the reset point (f). In the latter case,
the potential picture allows for a simple interpretation of the high
variability ($CV > 1$) under moderate noise intensity: short ISIs occur when
the particle moves directly from $v_r$ to $v_{th}$, while long ISIs correspond
to realizations in which the particle first performs an excursion to the
potential minimum and then hits threshold.}
\end{figure}

In this analogy, the ISI of the respective neuron model turns into the
first-passage time of the Brownian particle starting at the reset point
towards the threshold point. Depending on the model and, in particular, on the
value of $\mu$, the passage can occur already without noise ({\em tonic}
firing regime) or must be assisted by fluctuations ({\em noise-induced} firing
regime). The tonic regime can be most easily illustrated in case of the PIF
where the particle just slides down an inclined plane from reset to threshold
(see \bi{pot}a) , whereas noise is needed to reach the threshold whenever
there is a barrier present between reset and threshold (QIF for $\mu<0$, see
\bi{pot}c) or right at the threshold (LIF, $\mu<1$, see \bis{pot}(e) and
(f)). Note that the parameter $\mu$ has different meaning in the three
models. In the PIF it attains only positive values and sets merely the time
scale of the system. In the QIF it is a bifurcation parameter: at negative
$\mu$ the potential attains one minimum whereas for positive $\mu$ the
potential is a nonlinear but monotonic function. In the LIF, the bifurcation
from tonic to noise-induced firing takes place at $\mu=1$.  As we will see,
for the firing statistics of the LIF it is furthermore useful to distinguish
the case where $\mu<0$: here {\fett large values of the CV can be easily
interpreted in light of the specific properties of the potential} (see below).

\subsection{Measures}

The spike train is defined as a sum of
delta functions at the spiking times \citep{GerKis02}, i.e., the time instants
when the voltage reaches the threshold and the fire-and-reset rule is applied
(cf. \bi{spktr_fig}): 
\be
y(t)=\displaystyle\sum_j \delta(t-t_{j}).
\label{spike_train}
\ee 
In \e{spike_train}, $t_{j}$ stands for the instant when the $j$-th spike is
triggered.  \bi{spktr_fig} depicts the time evolution of the subthreshold
voltage as described by one of the models we address and the corresponding
spike train.  The time intervals $T_{j}=t_{j}-t_{j-1}$ between two immediately
subsequent spikes are the ISIs.

\begin{figure}[h!]
\centerline{\includegraphics[width=0.3\textwidth]{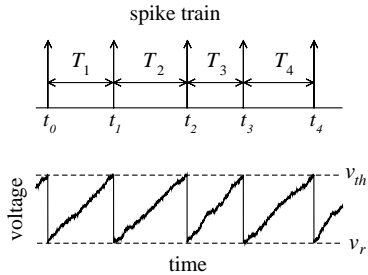}}
\caption{
\label{fig:spktr_fig}
Subthreshold voltage dynamics and corresponding spike train
as from a simulation of \e{if} and \e{f_pif}
(PIF) with parameters $\mu=0.9$ and $D=0.006$. {\fett Note the high degree 
of regularity of the spike train, characteristic of the PIF model 
under strong mean input with low noise intensity.}
}
\end{figure}

The spike trains considered here are stationary stochastic point processes.
The {\em firing rate} $r$ of such a process can be defined {\fett as the inverse mean ISI:}
%
\be
r=\frac{1}{\langle T \rangle}.
\label{rate_def}
\ee

The {\em coefficient of variation} ($CV$) of the ISI is defined as: 
\be
CV=\frac{\sqrt{\langle \Delta T^2 \rangle}}{\langle T \rangle},
\label{cv_def}
\ee
where $\langle \Delta T^2 \rangle=\lr{T^2} - \lr{T}^2$ is the variance of the
ISI distribution.  The CV can be regarded as the relative standard deviation
of the ISI. For comparison, a perfectly periodic spike train would have zero
CV while a Poissonian spike train possesses a CV of one.

As regards experimentally measured values of the rate and the CV, we note that
all nondimensional values of rates discussed in the following translate into
the former by $r_{real}=r/\tau$. For example, for a membrane time constant
$\tau=R_m C_m=10$ms nondimensional IF rates of $r=0.01,0.1,$ and $1$
correspond to firing rates of $r_{real}=1,10,$ and $100$ Hz. Note that the CV
(being a {\em relative} standard deviation) does not change upon a time
transformation, i.e. $CV_{real}=CV$.

\subsection{General form of the differential equations governing the contour lines}

Analytical formulas for the moments $\lr{T^n}$ of the first passage time from
$x_{-}$ to $x_{+}$ in an arbitrary potential $U(x)$ were derived by
\cite{PonAnd33}. Simplifications of these quadrature formulas as well as sum
formulas for specific cases have been put forward by many authors (for a
selection, see, for instance,
\citep{Hol76,Ric77,RicSac79,Tuc88,ColSan89,BulLow94,BulEls96,LinLSG02,LinLon03}).  The
first two moments determine the rate and CV, according to
Eqs.~(\ref{rate_def}) and (\ref{cv_def}).  For completeness, we write the
expressions for these moments here: 
\be 
\label{mean_general}
\langle T \rangle = \frac{1}{D}\int_{x_{-}}^{x_{+}}dx e^{U(x)/D}
\int_{-\infty}^{x}dy e^{-U(y)/D}
\ee
and
\be
\label{2mom_general}
\langle T^2 \rangle = \frac{2}{D^2}\int_{x_{-}}^{x_{+}}dx e^{U(x)/D} 
\int_{-\infty}^{x}dy e^{-U(y)/D} \times \int_{y}^{x^+} dz e^{U(z)/D}
\int_{-\infty}^{z}dv e^{-U(v)/D}.
\ee

In this paper we will study the rate and CV of the three models as functions
of the input parameters, $\mu$ and $D$.  In particular, we are interested in the curves for which
$F(D,\mu)=\mbox{const}$ ($F$ denotes either $r$ or CV), i.e., the contour
lines of the surfaces $F(D,\mu)$ over the $(D,\mu)$ parameter plane.
{\fett The contour lines can be obtained if one observes that $dF = 0$ along them. 
Since $dF=(\partial F / \partial D)dD + (\partial F / \partial \mu) d\mu$, 
one concludes that the following
differential equations for functions $\mu_F (D)$ or
$D_F (\mu)$ parametrize the contour lines:}
\be
\frac{d\mu_F }{dD}=-\frac{\partial F / \partial D}{ \partial F / \partial \mu},
\label{diffeq_contourline1}
\ee
\be
\frac{dD_F }{d\mu}=-\frac{\partial F / \partial \mu}{ \partial F / \partial D},
\label{diffeq_contourline2}
\ee
where $F\in \{r,CV\}$, provided that $\partial{F}/\partial{\mu}\neq 0$ and
$\partial{F}/\partial{D}\neq 0$, respectively.  
We note that these conditions are not necessarily satisfied in the whole
$(D, \mu)$ parameter space of the models we address.
For instance, for the PIF we have in fact $\partial{r}/\partial{D}=0$ 
for all valid pair $(D,\mu)$.
However, for the three models studied here, at any point
$(D,\mu)$ of parameter space at least one of these conditions is satisfied.

If for any pair $(r,CV)$ the respective contour lines
$\mu_r (D)$ and $\mu_{CV}(D)$ intersect at most once, then rate and CV
determine uniquely the parameters of the respective IF model.
In the following sections we will show that this is indeed the case
for the PIF, LIF, and QIF.


\section{Perfect integrate \& fire neuron}\label{sec:pif}

The mean and variance of the ISI are given by \citep{Hol76,Tuc88,  BulLow94, Bur06}:
\begin{equation}
\langle T \rangle = \frac{v_{th} - v_r }{\mu}, \hspace{5mm} \langle \Delta T^2 \rangle =
\frac{2D(v_{th}-v_r )}{\mu^3}.
\end{equation}
We stress that $\mu >0$ for the PIF; otherwise all moments of the ISI
diverge. For this model the expressions for rate and CV are quite simple:
\begin{equation}
r = \frac{\mu}{v_{th} - v_r}, \hspace{5mm} CV^2 =
\frac{2D}{\mu (v_{th}-v_r )}. \label{pif1}
\end{equation}
Moreover, the contour lines for the rate and the CV
can be explicitely calculated (without resorting to the differential equations
\e{diffeq_contourline1} and \e{diffeq_contourline2}):
\be \mu_{r_0 }(D)=r_0 (v_{th}-v_r),\;\;\;\; \mu_{CV_0
}(D)=\frac{2D}{(v_{th}-v_r )CV_{0}^{2}}.
\label{pif4}
\ee 
We briefly review the behavior of rate and CV as functions of $\mu$ and $D$
and then show that rate and CV uniquely fix the system's parameters.
\subsection{Rate and CV and their contour lines in the ($D,\mu$) plane for the PIF}
\begin{figure}[h!]
\centerline{\includegraphics[width=0.35\textwidth]{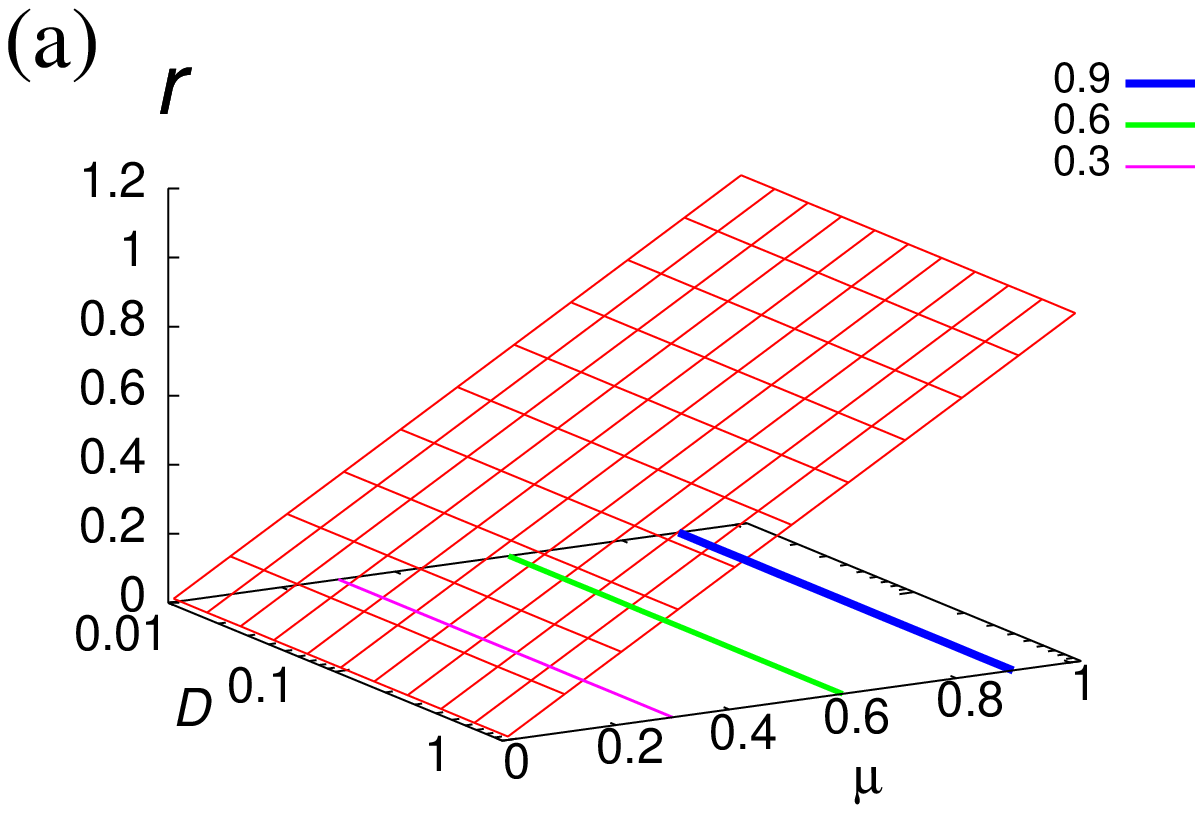}}
\centerline{\includegraphics[width=0.35\textwidth]{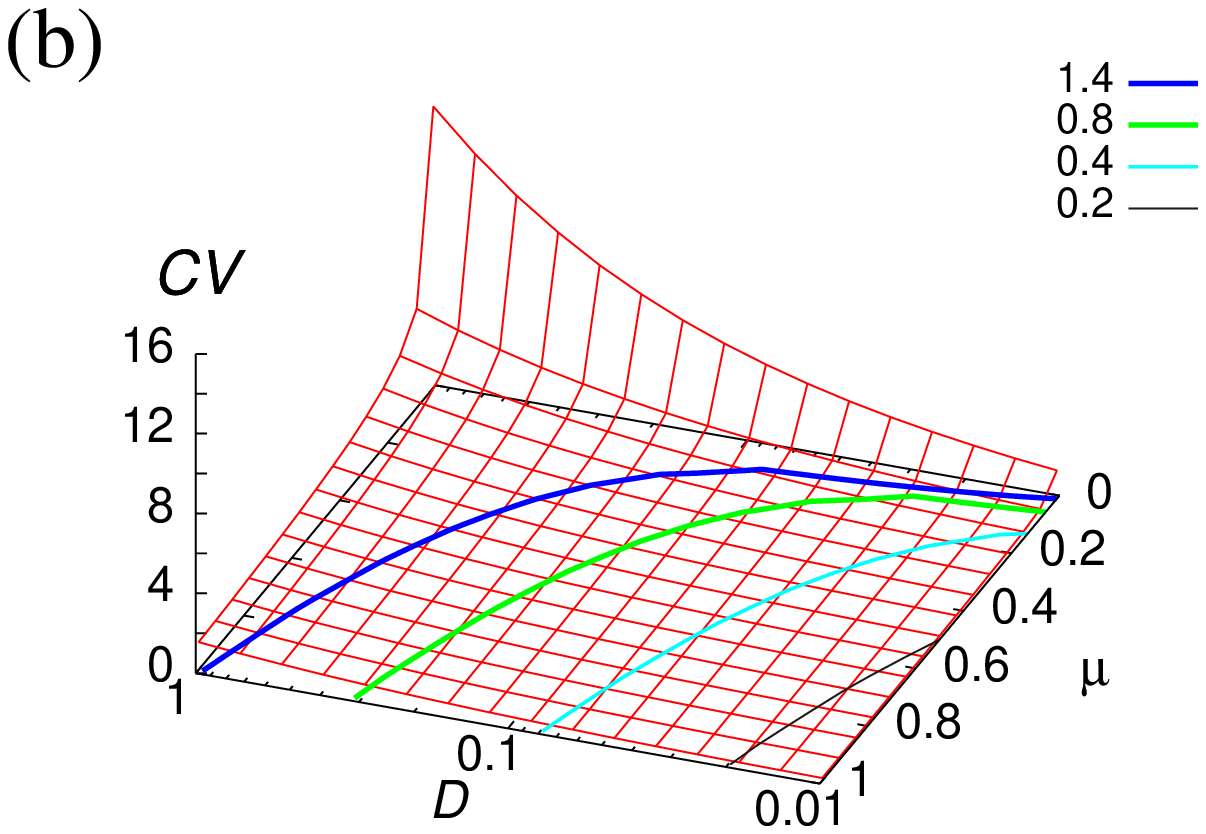}}
\centerline{\includegraphics[width=0.35\textwidth]{contoursPIF.eps}}
\caption{
\label{fig:rcvPIF}
Rate (a) and CV (b) as a function of the parameters $\mu$ and $D$ for the PIF
model. The rate is a simple increasing function of the mean input and does not
depend on the noise intensity. The CV is proportional to $\sqrt{D/\mu}$
(cf. \e{pif1}). In (c) we show contour lines corresponding to three different
values of the rate (dashed lines) and to four different values of the CV
(solid lines with symbols). If for a given pair of rate and CV there is only
one pair of input parameters $(\mu,D)$, there should be only one intersection
point for two specific (dashed and solid) contour lines. The uniqueness of
this correspondence can be explicitely shown for the PIF (see \e{pif2} and \e{pif3}).
}
\end{figure}

The rate and CV are shown in Figs. \ref{fig:rcvPIF}(a) and \ref{fig:rcvPIF}(b) as
functions of the parameters $\mu$ and $D$.  The rate is a linear function of
$\mu$ and, remarkably, does not display any dependence on $D$.  This is a
unique property of the PIF model.  The $CV$ depends linearly on
$\sqrt{D/\mu}$, and can therefore attain values in the whole range $0<CV<\infty$.

\bi{rcvPIF}(c) shows contour lines for different rates and CVs,
which are for both measures just straight lines. Generally, the variability of 
the PIF's spike train increases by decreasing the mean
input and increasing the noise intensity, which is quite intuitive.

\subsection{Uniqueness of the model parameters for a given rate and CV for the PIF} 
In the fairly simple case of the PIF,
\e{pif1} can be readily
inverted to yield $\mu$ and $D$ as a function of
rate and CV:
\begin{equation}
\mu=r(v_{th}-v_r), \label{pif2}
\ee
\be
D=\frac{r(v_{th}-v_r)^2 CV^2} {2}. \label{pif3}
\end{equation}
Eqs. (\ref{pif2}) and (\ref{pif3}) define a mapping $(r,CV) \mapsto
(D,\mu)$, implying  that for any pair $(r,CV)$ there exists one and
only one pair $(D,\mu)$.


\section{Leaky integrate \& fire neuron}\label{sec:lif}

For the LIF, the mean and variance of the ISIs are \citep{Ric77,Tuc88,LinLSG02}:
\begin{equation}
\langle T \rangle = \sqrt{\pi}\int_{a}^{b}dy e^{y^2} \mbox{ erfc}(y),\label{meanISIlif}
\end{equation}
\begin{equation}
\langle \Delta T^2 \rangle =
2\pi \int_{a}^{b}dz e^{z^2} \int_{z}^{\infty} dy e^{y^2} \mbox{erfc}^2(y), \label{lif1}
\end{equation}
where 
\be
a=(\mu-v_{th})/\sqrt{2D}
\mbox{    and    } 
b=(\mu-v_r)/\sqrt{2D}.
\label{a&b}
\ee

From these expressions and the general relations \es{diffeq_contourline1} and
 \e{diffeq_contourline2}, one can derive the differential equations that
govern the contour lines as follows:
\ba 
\frac{d\mu_{r}}{dD}&=& \frac{b-a}{v_{th}-v_{r}}\left(\frac{be^{b^2} \mbox{
erfc}(b)-ae^{a^2}\mbox{ erfc}(a)}{e^{b^2}\mbox{ erfc}(b)-e^{a^2}\mbox{
erfc}(a)}\right), \label{dmurdD_lif} \\ 
\frac{d\mu_{CV}}{dD}&=&
\left(\frac{b-a}{v_{th}-v_r }\right)\left[a\left(1-\mathbb{F}(a,b)\right)^{-1}
+b\left(1-\frac{1}{\mathbb{F}(a,b)}\right)^{-1}\right], 
\\ 
\mathbb{F}(a,b)&=&
\frac{\int_{a}^{b}dx e^{x^2}\mbox{erfc}(x) e^{b^2}\int_{b}^{\infty}dy
e^{y^2}\mbox{erfc}^2(y) -2\int_{a}^{b}dze^{z^2}\int_{z}^{\infty}dy
e^{y^2}\mbox{erfc}^2(y) e^{b^2}\mbox{erfc}(b)} {\int_{a}^{b}dx
e^{x^2}\mbox{erfc}(x) e^{a^2}\int_{a}^{\infty}dy e^{y^2}\mbox{erfc}^2(y)
-2\int_{a}^{b}dze^{z^2}\int_{z}^{\infty}dy e^{y^2}\mbox{erfc}^2(y)
e^{a^2}\mbox{erfc}(a)}, 
\\ 
\frac{dD_{CV}}{d\mu}&=& \left[ \frac{d\mu_{CV}}{dD}
\right]^{-1}.  
\ea 
We will first recall some properties of rate and CV, most but not all of which
have been already discussed elsewhere \citep{PakTan01,LinLSG02}.

\subsection{Rate and CV and their contour lines in the ($D,\mu$) plane for the LIF}
Rate and CV as functions of $\mu$ and $D$ are shown in \bi{rcvLIF}.  As seen
in \bi{rcvLIF}(a), the rate is an increasing function of $\mu$ for fixed $D$
and an increasing function of $D$ for fixed $\mu$.  In the zero-noise limit,
the rate is zero for $\mu < v_{th}$ and increases monotonically for 
$\mu > v_{th}$ according to the well-known rate of a deterministic LIF model
$r^{-1}=\ln[(\mu-v_r)/(\mu-v_{th})]$ \cite{GerKis02}. 

The behavior of the CV is much richer (see \bi{rcvLIF}(b)).  In particular,
the LIF model displays {\it coherence resonance} (CR)
\citep{PakTan01,LinLSG02}: for fixed $\mu<v_{th}$, the CV exhibits a minimum at
a finite value of $D$.  Coherence resonance thus corresponds to the phenomenon
by which noise has the counter-intuitive effect of increasing the regularity
of the spike train.  For the LIF, a pronounced CR is observed for a mean input
$\mu$ close to but smaller than the threshold ($\mu \lesssim v_{th}$).

The CV for the LIF can exceed unity (this feature is also displayed by the PIF
but not by the QIF).  Loosely speaking, such a regime corresponds to a firing
activity more irregular than in the Poissonian regime ($CV=1$).  This high
variability is associated to short ISIs occurring relatively frequently, but
long ISIs being also likely.  When $\mu<0$, a simple interpretation can be
made in terms of the Brownian particle in a parabolic potential.  As shown in
\bi{pot}(f), in this case, both $v_r$ and $v_{th}$ are larger than the value of
$v$ at which the potential attains its minimum.  The short ISIs then
correspond to the cases when the particle heads directly from $v_r$ to
$v_{th}$, while the long ones correspond to the particle first going to the
minimum of the potential and then performing its excursion to $v_{th}$.

Independently of $D$, if $\mu \rightarrow -\infty$ the firing becomes
Poissonian ($CV=1$).  In the opposite limit of $\mu \to \infty$, the firing is
perfectly regular ($CV=0$).  At least for large noise intensity, the CV
exceeds unity, as discussed above.  Therefore, for fixed and sufficiently
large value of the noise intensity we observe a maximum\footnote{The existence
of this maximum was pointed out to the authors by Tilo Schwalger, {\it
Max-Planck-Institut f\"ur Physik Komplexer Systeme (Dresden)}.} of the CV with
respect to $\mu$.  This is an interesting feature of the LIF model - not
shared by PIF or QIF - which to our knowledge has not been described so far.
{\fett It implies that, given a fixed level of the input fluctuations, the
degree of irregularity of the spike train is maximal ({\it maximized
incoherence}) for a finite value of the mean input.  We note that another type
of maximized incoherence has been described for the LIF with an additional
absolute refractory period by \cite{LinLSG02}. In that case, the CV achieves a
maximal value for a finite noise intensity when the mean input is fixed.}

\begin{figure}[h!]
\centerline{\includegraphics[width=0.35\textwidth]{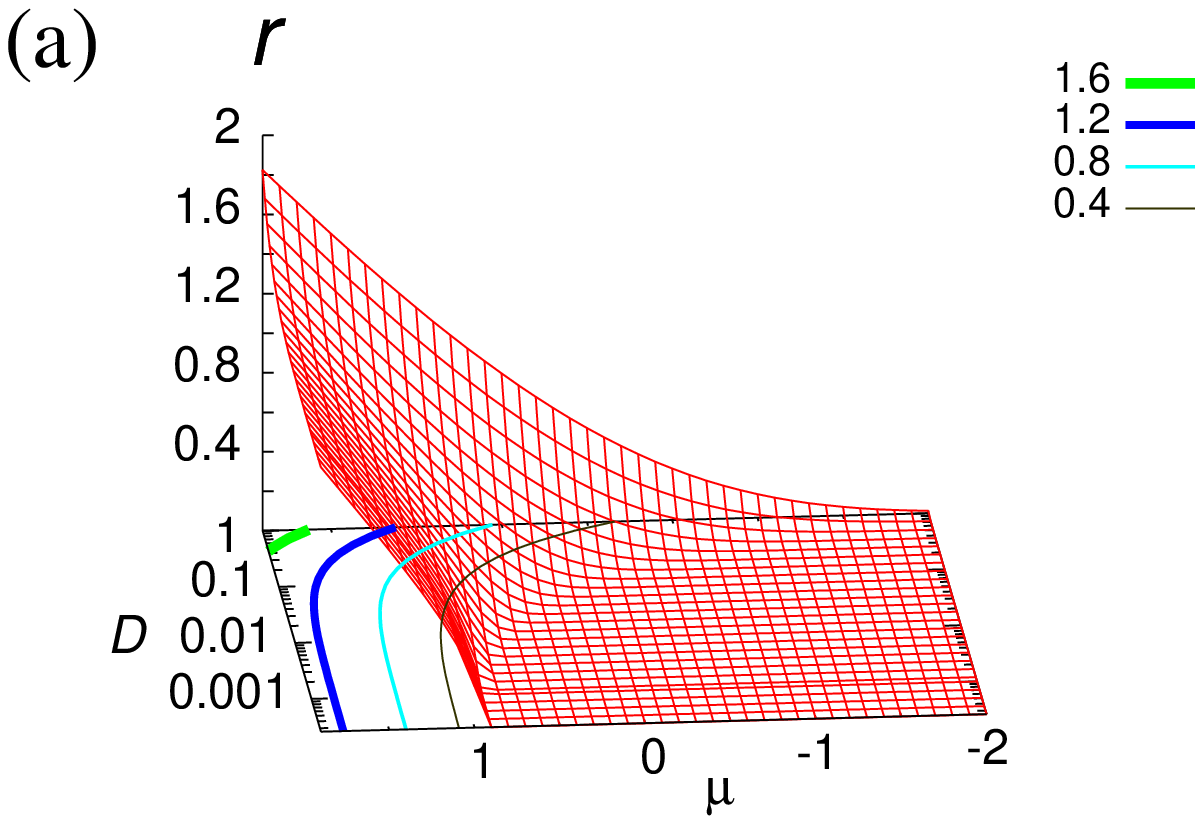}}
\centerline{\includegraphics[width=0.35\textwidth]{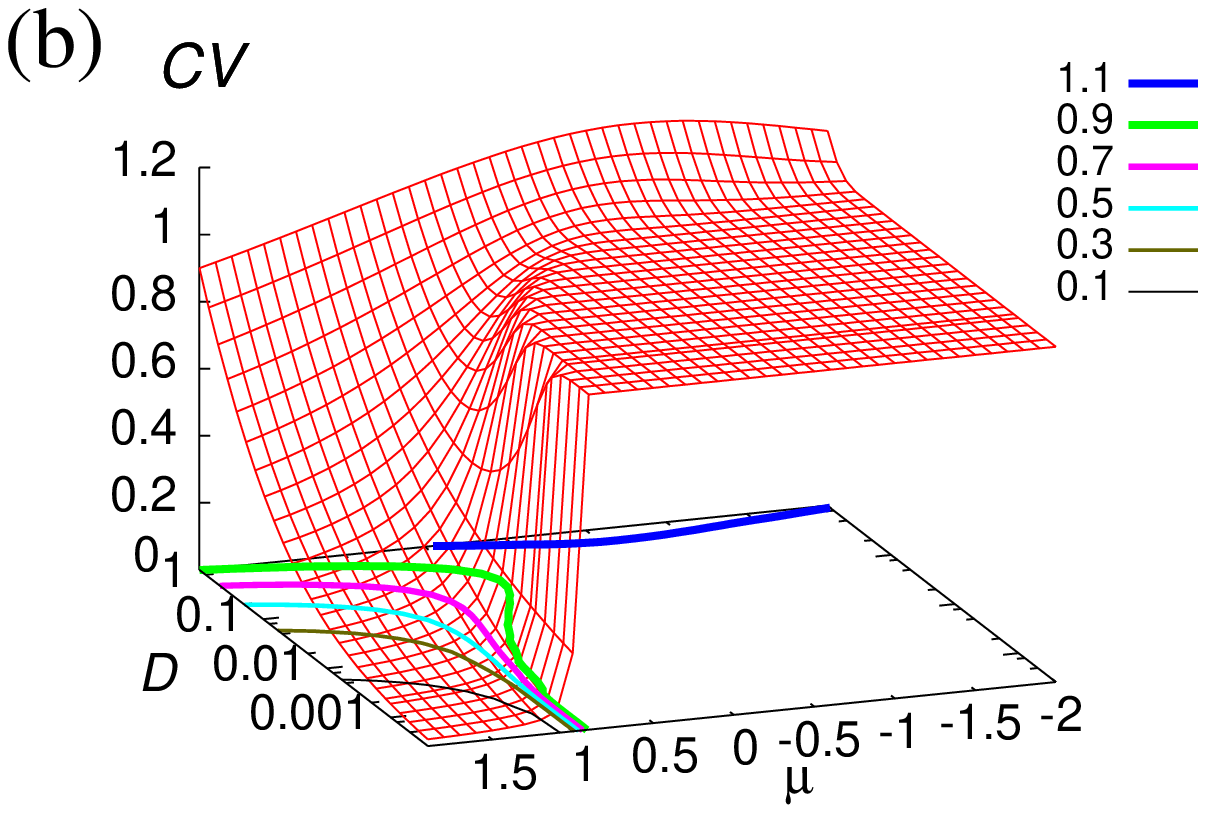}}
\centerline{\includegraphics[width=0.35\textwidth]{contoursLIF_withPoisson.eps}}
\caption{
\label{fig:rcvLIF}
Rate (a) and CV (b) as a function of the parameters $\mu$ and $D$ for the LIF
model.  In (c) we show contour lines for three values of the rate (dashed
lines) and four values of the CV (solid lines with symbols). {\fett Note the
nonmonotonic character of the CV's contour lines $\mu_{CV_0}(D)$.}  The
rectangular region marked in (c) is magnified in (d).  {\fett Also plotted in
(d) is one curve (brown thick line) of the type $a=\mbox{const}$ [cf. \e{a&b}], corresponding to
a contour line for the ISI statistics in the Poissonian firing limit. Contour
lines in this limit for the rate ($r=0.01$, black dashed line) and CV
($CV=0.95$, magenta triangles) are very close to this line and, consequently,
to each other.} }
\end{figure}

As also shown in \bi{rcvLIF}(b), the contour lines for the CV display
nonmonotonicities with respect to both parameters.  The contour lines at which
the CV is smaller than 1 display nonmonotonic behavior with respect to $D$,
whereas the ones corresponding to CV larger than 1 are nonmonotonic functions
of $\mu$ (see, for instance, the contour line $CV=1.1$).
\bi{rcvLIF}(c) shows contour lines for rate and CV in a range of physiological
interest.

We also observe in \bi{rcvLIF}(d) that the contour lines of rate and CV are
very close to each other in the region of small $D$ and $\mu<v_{th}$ (see
especially contour lines $r=0.01$ and $CV=0.95$).  {\fett The reason is that
in this region we approach the Poissonian firing limit, in which case the
dynamics of the model depends on the threshold value --- hence on $a$ --- but
not on the reset value --- hence not on $b$ (see e.g.  \cite{DitLan05} and
references therein). For a small but finite value of $D$ and $\mu<v_{th}$, the
dependence on the parameter $b$ that distinguishes the contour lines of rate
and CV is very weak which is the reason why these contour lines will be very
close to each other. Assuming an exclusive dependence on $a$, we obtain a
universal contour line for {\em all} ISI statistics (this holds strictly true
only in the Poissonian limit) by setting $a=$const .  One such line (with
$a=-2.1$ extracted from the rate curve for $r=0.01$ at $D=0.001$) is shown in
\bi{rcvLIF}(d): indeed, the corresponding contour lines of rate and CV are very
close to this line.}

Thus, as the firing regime approaches the Poissonian limit, the actual
determination of the intersections of the contour lines becomes a practically
more difficult task.  Also it becomes less clear whether there is only
intersection point or not.  In view of this particular (numerical)
uncertainty, but also in view of the nonmonotonic behavior of the contour
lines $\mu_{CV}$ as functions of $D$ (for $\mu<v_{th}$) and $\mu$ (for strong
noise), it is desirable to gain certainty about whether rate and CV uniquely
determine $D$ and $\mu$ in the LIF model.

\subsection{Uniqueness of the model parameters for a given rate and CV for the
LIF} 

Our strategy to show that the model parameters are uniquely determined for a
given rate and CV comprises two steps.  First, we demonstrate that each
contour line for the rate is unique.  Second, we prove that the CV is a
monotonic function along any rate contour line. The second step can be
simplified by noting that the CV is the ratio between the square root of the
{\it variance} $\sigma^2 =\la \Delta T^2 \ra$ and the mean $\lr{T}$. Since the
mean is invariant in any contour line for the rate, it suffices to show that
the $\sigma^2$ is a monotonic function along any such contour line.  In other
words, it suffices to show that the directional derivative of $\sigma^2 $
along the tangent of the contour line for the rate is strictly positive
\footnote{The basic idea for this step in the proof is due to Dr. Jochen
Br\"ocker, {\it Max-Planck-Institut f\"ur Physik Komplexer Systeme
(Dresden)}.}.

\subsubsection{Uniqueness of contour lines for the rate}

Let us prove that the contour line for a specific value of the rate
$r(D,\mu)=r_0$ is one single connected curve.  The proof comprises three
steps: First, for any point $(D,\mu) \in \mathbb{R}^+ \times \mathbb{R}$ we
can locally construct a contour line $\mu_{r_0}(D)$ 
parametrized by 
$D$ such that $r(D,\mu_{r_0}(D))=r_0$.  This is possible locally by virtue of
the implicit function theorem since, as shown in Appendix \ref{app:rate_props},
$\partial{r}/\partial{\mu}>0$ for all $(D,\mu)\in \mathbb{R}^+ \times
\mathbb{R}$.  Of course, the specific value of the rate, $r_0$, will depend on
the point $(D,\mu)$.  Second, we can extend $\mu_{r_0}(D)$ to the whole domain
$D\in \mathbb{R}^+$ by connecting neighborhoods of this domain.  This could be
made impossible if $\mu_{r_0}(D)$ diverges at finite D. We rule out this
possibility by noting that it is not consistent with the limit values of
the rate at $\mu=\pm\infty$.  Indeed, as we show in Appendix \ref{app:rate_props}, the
limit of the rate is zero (for $\mu \to -\infty$) or $\infty$ (for
$\mu\to\infty$), values which are not attained by the rate for any positive $D$ and
finite $\mu$. Hence no contour line starting within the domain can approach
the boundary $\pm \infty$ at finite $D$ and thus there cannot be a divergence
of the contour line at finite $D$. Hence $\mu_{r_0}(D)$ will describe a single
connected line for the whole domain $D\in \mathbb{R}^+$.  Third, we show that
the graph of $\mu_{r_0}(D)$ contains all points $(D,\mu)$ with
$r(D,\mu)=r_{0}$.  In fact, if a point $(D^*,\mu^*)$ not belonging to the
graph of $\mu_{r_0}(D)$ exists such that $r(D^*,\mu^*)=r_{0}$, then the
condition $\partial{r}/\partial{\mu}>0$ must necessarily be violated along the
vertical line $D=D^*$.  This completes the proof that the contour lines for rate are
single (connected) curves.

\subsubsection{Proof that $\sigma^2$ is a monotonic function along
the rate contour lines}
Along a contour line of the rate, the mean ISI is fixed by definition and thus
the CV can only vary due to changes in the variance $\sigma^2$. Thus if we
show that the variance increases monotonically as we move along the contour line
in the direction of increasing $D$, we will have also shown that the CV
increases monotonically if we move along the contour line in this direction.

The monotonicity of \var along the rate contour lines is expressed by
\be
\nabla \sigma^2 \cdot {\bf v_t} > 0,
\label{thesisLif}
\ee
where ${\bf v_t}$ is a vector which is tangent to these contour lines and
$\nabla$ denotes the gradient in $(D,\mu)$ space.  In order to show that
\e{thesisLif} holds, let us first determine ${\bf v_t}$.  Along the
rate contour lines, the differential equation
\e{diffeq_contourline1} with $F=r$ holds true; its right-hand side is needed
for an expression of the tangent vector of the rate contour lines appearing in
\e{thesisLif}:
\be
{\bf v_t}={\bf e}_D + \frac{d\mu_r}{dD}{\bf e}_\mu
\ee
where ${\bf e}_D$ and ${\bf e}_\mu$ are the respective unit vectors. The
relation to be shown, \e{thesisLif}, thus corresponds to
\be
\frac{\partial \sigma^2}{\partial D} + \frac{\partial \sigma^2}{\partial \mu} \frac{d\mu_r }{dD} >0.
\label{thesisLif2}
\ee

It is much simpler to express the derivatives on the left hand side of
Eq.(\ref{thesisLif2}) in terms of coordinates $a$ and $b$ rather than $D$ and
$\mu$. Using Eqs. (\ref{lif1}) and (\ref{a&b}), we obtain
\be
\frac{\partial \sigma^2}{\partial D}=
\frac{2\pi(b-a)^2}{(v_{th}-v_r )^2} \left( -be^{b^2}\int_{b}^{\infty}dy e^{y^2}\mbox{erfc}^2(y) + a
e^{a^2}\int_{a}^{\infty} dy e^{y^2}\mbox{erfc}^2(y)
\right)
\label{ds2dD}
\ee
and
\be
\frac{\partial \sigma^2}{\partial \mu}=
\frac{2\pi (b-a)}{v_{th}-v_r}
\left(
e^{b^2}\int_{b}^{\infty}dy e^{y^2}\mbox{erfc}^2(y) 
-e^{a^2}\int_{a}^{\infty} dy e^{y^2}\mbox{erfc}^2(y)
\right).
\label{ds2dmu}
\ee Inserting Eqs. (\ref{dmurdD_lif}, \ref{ds2dD}-\ref{ds2dmu}) into
Eq.(\ref{thesisLif2}), and performing straightforward algebra, we write the
latter as: 
\be \frac{2\pi(b-a)^3 e^{a^2 +b^2
}\mbox{erfc}(a)\mbox{erfc}(b)}{(v_{th}-v_{r})^2 (e^{a^2}\mbox{erfc}(a)-
e^{b^2}\mbox{erfc}(b))}\left(\int_{a}^{\infty}dy
e^{y^2}\frac{\mbox{erfc}^2(y)}{\mbox{erfc}(a)} -\int_{b}^{\infty}dy
e^{y^2}\frac{\mbox{erfc}^2(y)}{\mbox{erfc}(b)} \right) >0.
\label{thesisLif3}
\ee 

This inequality holds true for all $a<b$ since the two functions $e^{a^2}
\mbox{erfc}(a)$ and $\int_a^\infty dx \;\mbox{erfc}^2(x)/\mbox{erfc}(a)$
(differences of which appear in \e{thesisLif3}) are monotonically decreasing
functions of $a$ as we prove in the Appendix \ref{app:ineq}. We have thus
proven \e{thesisLif2} and hence we have shown that the variance of the ISI
always increases as we go along the constant rate contour line in the
direction of increasing noise intensity. This completes our proof of the
uniqueness of parameters determined by prescribed values of the rate and the
CV.


\section{Quadratic integrate \& fire neuron}
\label{sec:qif}

For the QIF, one has \citep{LinLon03}:
\be
\langle T \rangle = \left( \frac{9}{D} \right)^{1/3} I(\alpha)\mbox{,    }\hspace{4mm}  I(\alpha)=\int\limits_{-\infty}^{\infty}
dx\; e^{-\alpha x-x^3} \int\limits_{-\infty}^{x} dy\; e^{\alpha y+y^3}, \label{meanqif}
\ee
\be
\langle \Delta T^2 \rangle =
\left(\frac{9}{D}\right)^{2/3} \int_{-\infty}^{\infty}dx
e^{-\alpha x -x^3}
\int_{x}^{\infty} dy e^{-\alpha y-y^3}
\left[ \int_{-\infty}^x dz e^{\alpha z+z^3 } \right]^2 , \label{varqif}
\end{equation}
\be
\alpha=\left( \frac{3}{D^2 } \right)^{1/3}\mu. \label{alpha}
\ee
%
For this model, the following scaling relations \citep{LinLon03}
facilitate the determination of the contour lines in parameter
space for rate and CV:
\be
r(\mu,D)=\sqrt{|\mu |} r(\frac{\mu}{|\mu|},|\mu |^{-3/2}D),
\label{scaling_rate}
\ee
\be
CV(\mu,D)=CV(\frac{\mu}{|\mu|},| \mu |^{-3/2}D).
\label{scaling_cv}
\ee
The scaling relation \e{scaling_cv}, 
together with the monotonicity of the CV for $\mu=\pm1$ (see \cite{LinLon03}), 
implies that (for $\mu=\pm1$) a certain value of the CV (say, $CV_0$) determines uniquely the noise intensity D which we call $\bd$.
With this observation, the 
contour lines for the CV for arbitrary $\mu$ can be explicitely written:
\be
\mu_{cv}(D)=\frac{\mu}{|\mu|} \left(\frac{D}{\bd}\right)^{2/3}.  
\label{mu_cv_of_D_qif}
\ee 

For the rate, $\bd$ can be regarded as a parameter of the curve $\mu_r(D)$:
from the above definition of $\bd$ and from the scaling relation for the rate \e{scaling_rate} we
can infer that
\be
\label{mu_D_barD}
\mu(\bd)=\frac{\mu}{|\mu|}
\left(\frac{r_0}{r(\frac{\mu}{|\mu|},\bd)}\right)^2,\;\;\; D(\bd)=
\bd\left(\frac{r_0}{r(\frac{\mu}{|\mu|},\bd)}\right)^3
\label{contour_rate_qif}
\ee
describe all the points on the curve $\mu_r(D)$ which we get by varying
$\bd$. 

We will now recall some properties of rate and CV 
which have been already discussed by \cite{LinLon03}.

\subsection{Rate and CV and their contour lines in the ($D,\mu$) plane for the
QIF}

Rate and CV as a function of $\mu$ and $D$ are shown in \bi{rcvQIF}.  The
behavior of the rate is similar to the case of the LIF.  Here again it is a
monotonically increasing function of $\mu$ for fixed $D$ and monotonically
increasing function of $D$ for fixed $\mu$.  A noticeable difference arises in
the zero-noise limit and close to the bifurcation at $\mu=0$: the rate for the
QIF is also strictly zero if $\mu\leq0$ but, differently from what is observed
for the LIF, it increases proportionally to the square root of  $\mu$ for
small positive $\mu$.

In clear contrast to the LIF and PIF, the CV for the QIF is bounded in the
interval $0<CV<1$.  Moreover, here the CV is a strictly monotonic function of
both $D$ and $\mu$. For fixed $D$, it decreases with increasing $\mu$ and, for
fixed positive (negative) $\mu$, it increases (decreases) with increasing $D$.
Therefore, coherence resonance does not occur for this model.
 
\bi{rcvQIF}c shows some contour lines for rate and CV in the physiologically
relevant region of the parameter space.  We emphasize that for the QIF the
contour lines of the CV are monotonic functions of both $D$ and $\mu$.

{\fett As shown in \bi{rcvQIF}d, the contour lines of rate and CV are in close
  vicinity in the region of small $D$ and $\mu<0$, analogously to what is
  observed for the LIF.
In this Poissonian firing limit, the dynamics of the QIF depends most strongly
on the ratio of potential barrier and noise intensity \citep{LinLon03} which
can be interpreted as a rescaled barrier
\be
\label{const_barrier}
\Delta u= \frac{4|\mu|^{3/2}}{3D}.
\label{aqif}
\ee
For instance, the firing rate in the Poissonian limit (corresponding to the
Kramers rate out of the cubic potential well) depends on this barrier
exponentially and has only a mild additional dependence on $\mu$ via a
prefactor \citep{LinLon03}
\be
\label{rate_smallD_qif}
r=\sqrt{|\mu|}/\pi e^{-\Delta u}, \;\;\; \Delta u\gg 1. 
\ee 
The line 
$\Delta u=$const is indeed close to both the contour lines of rate and of CV
in the Poissonian limit (cf. brown line in \bi{rcvQIF}d).  Note that
\e{rate_smallD_qif} offers another, more accurate approximation, to the
contour line of the rate in the Poissonian firing regime. This line is,
however, very close to the line obtained from a constant barrier $\Delta
u=$const.}

\begin{figure}[h!]
\centerline{\includegraphics[width=0.35\textwidth]{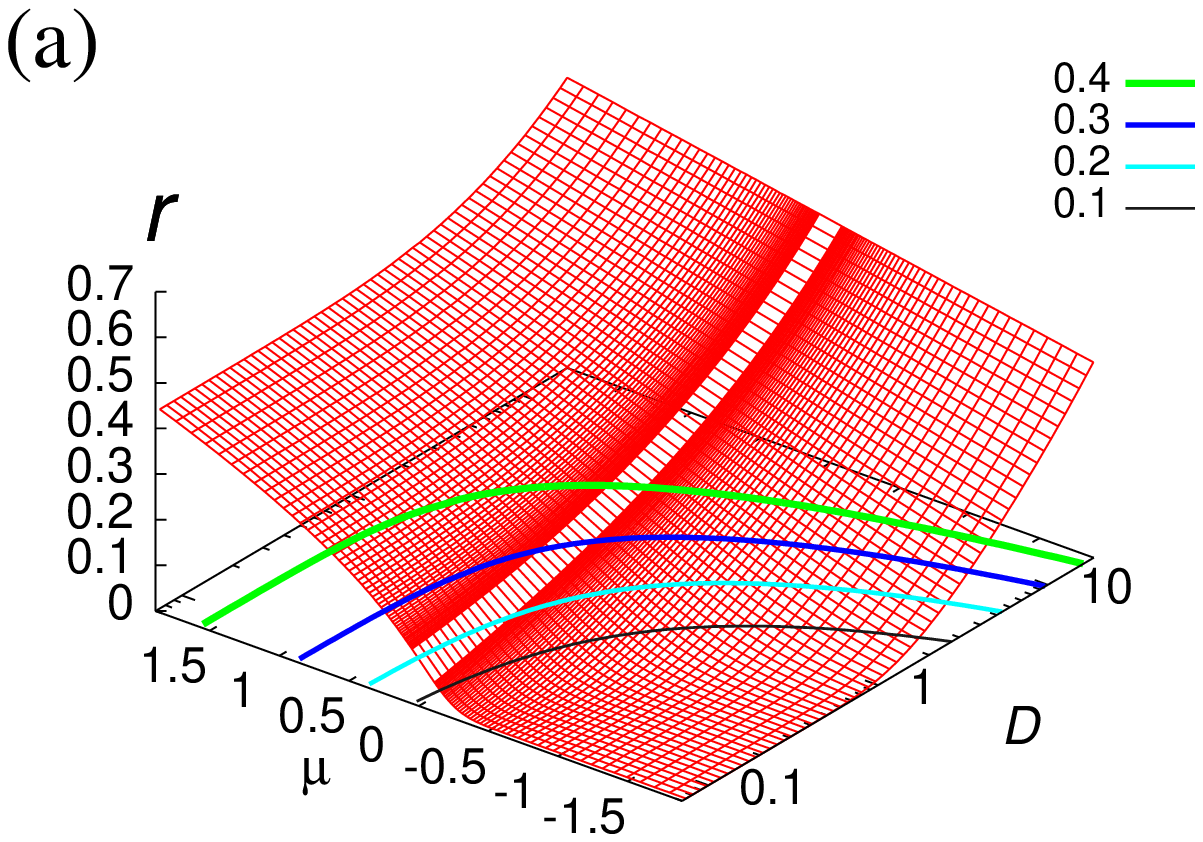}}
\centerline{\includegraphics[width=0.35\textwidth]{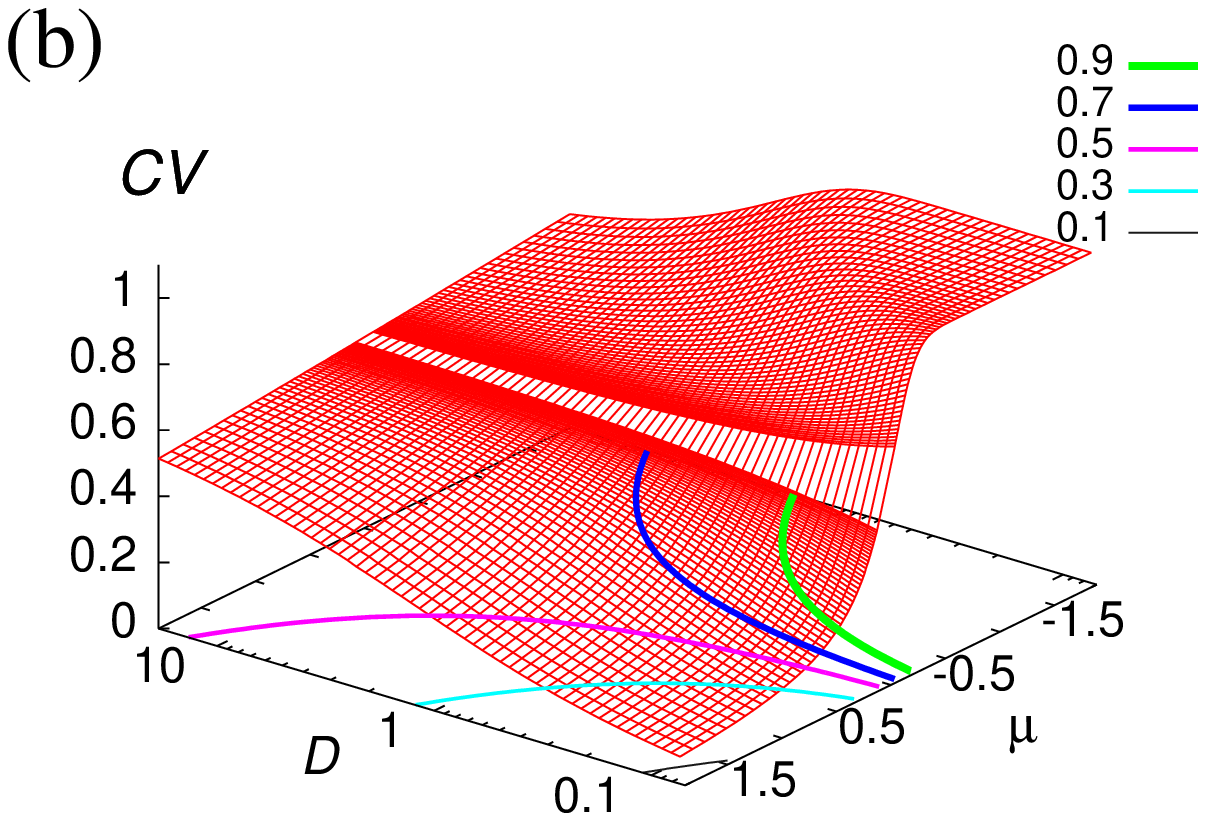}}
\centerline{\includegraphics[width=0.35\textwidth]{contoursQIF_withPoisson.eps}}
\caption{
\label{fig:rcvQIF}
Rate (a) and CV (b) as a function of the parameters $\mu$ and $D$ for the QIF
model.  In (c) we show contour lines for three values of the rate (dashed
lines) and four values of the CV (solid lines with symbols) as indicated. The
rectangular region in (c) is magnified in (d).  {\fett In (d) we also plot the
one curve (brown thick line) governing all ISI statistics in the Poissonian
regime with $\Delta u(\mu,D)=2.84$ (corresponding to the value at $D=0.1,
\mu=-0.36$) resulting in $\mu=-(0.75 D \Delta u)^{2/3}$ (cf. the discussion
around \e{const_barrier}).}  The more sparse grid in the region close to
$\mu=0$ in (a, b) is due to the way we generated the points and does not
reflect any property of the surfaces $r(D,\mu)$ and $CV(D,\mu)$.  }
\end{figure}

\subsection{Uniqueness of the model parameters for a given rate and CV for the QIF} 
We will consider as given that the $CV(1,\bd)$ ($CV(-1,\bd)$) is a
monotonically increasing (decreasing) function of $\bd$; this was previously
demonstrated by limit cases and by numerical evaluation of the integrals
\citep{LinLon03}. From these properties we can conclude: for negative
(positive) $\mu$, decreasing (increasing) the CV from 1 (0) to $3^{-1/2}$, the
parameter $\bd$ changes monotonically from 0 to infinity implying that each CV
between 0 and 1 has one unique contour line parametrized by the sign of $\mu$
and the value of $\bd$ (see Eq.(\ref{mu_cv_of_D_qif})). The value $\mu=0$ is a
special case where the CV attains exactly the value at the boundary between
the regimes $\mu < 0$ and $\mu>0$, namely, CV=$3^{-1/2}$ \citep{SigHor89}.

If we can show that the rate or equivalently the mean ISI changes
monotonically along the contour lines of the CV, then there is at most one
intersection for a given pair of CV and rate and thus the mapping of rate and
CV to $\mu$ and $D$ is unique.  Note that although our argument is similar to
the one used for the LIF, we will consider the change in the mean ISI along
the curve of constant CV and not the change in CV (or variance) along a curve
of constant rate as we did for the LIF.

The directional derivative of the mean ISI along the CV contour line reads
\be
\nabla \lr{T} \cdot {\bf v_t}=\frac{\partial \lr{T}}{\partial
  D}+\frac{\partial \lr{T}}{\partial \mu}\frac{d \mu_{CV}}{d D}.
\ee
From \e{mu_cv_of_D_qif}, we obtain:
\be
\label{diff_cv}
\frac{d\mu_{cv}}{dD}=\frac{\mu}{|\mu|} \frac{2}{3}D^{-1/3}\bd^{-2/3}. 
\ee 
Using this expression and Eqs. (\ref{meanqif}) and (\ref{alpha}), one can rewrite the directional derivative as follows:
\ba
\nabla \lr{T} \cdot {\bf v_t}&=&-\frac{1}{3D}\lr{T}-\frac{2}{D^{2}} \frac{\partial I(\alpha)}{\partial \alpha}\left[\mu-\frac{\mu}{|\mu|} \left(\frac{D}{\bd }\right)^{2/3}\right]
\ea
where we have used the auxiliary function $I(\alpha)$ from \e{meanqif}.  The
expression in the brackets vanishes by definition since it equals
$\mu-\mu_{CV}$ according to \e{mu_cv_of_D_qif}.  Hence we find that the last
term is truly zero and thus the directional derivative of the mean ISI along
the contour lines of the CV is negative throughout the $(\mu,D)$ plane
\ba
\nabla \lr{T} \cdot {\bf v_t}&=&-\frac{1}{3D}\lr{T}<0.
\ea
We have thus shown that (i) for each CV between 0 and 1 there exists exactly
one contour line and (ii) the mean ISI decreases always as we go along these
contour lines in direction of increasing noise intensity. Hence, each mean ISI
is at most represented once on a contour line of the CV and thus for each pair
of rate and CV values there is at most one pair $(\mu,D)$.

\section{Conclusions}\label{sec:conc}

To summarize, we have reviewed the behavior of rate and CV as functions of the
input parameters for three different IF models. As the central result of our
paper, we have shown that these statistics uniquely determine the input
parameters for the models studied. This sets a framework for systematic
comparison of these models: they can be compared on an equal footing when their
parameters are tuned so as to yield the same rate and CV. Reports on these
comparisons will be published elsewhere.

It is tempting to consider the general IF model with white noise input: do
rate and CV determine the input parameters for an arbitrary nonlinear function
$f_{\mbox{\tiny{model}}}(v)$ or equivalently for an arbitrary nonlinear
potential $U(v)$?  Unfortunately, so far no general procedure to show the
uniqueness of input parameters is known. What we showed in this paper relied
on model-specific properties of the ISI moments for the PIF, LIF, and
QIF. Approaches to the uniqueness problem based on the general formulas for
the moments of the first-passage time \e{mean_general} and \e{2mom_general}
could lead to conditions on the potential $U(v)$ and the reset and threshold
values of the IF model; however, we have not made any progress in this
direction yet. We note that it may still be worth the effort to prove the
uniqueness of parameter values for a given rate and CV for other specific
neuron models of the IF type. One such a case is the exponential IF model
\citep{FouHan03}, which has been successfully used to describe pyramidal
neurons activity \citep{BadLef08}.  

Another and more complicated open problem is to check whether the uniqueness
of parameters also holds for more complex models, i.e.  multidimensional
extensions of the simple IF model as those taking adaptation \cite{LiuWan01}
and threshold fatigue \cite{ChaLon00,ChaLin04,LinCha05}, subthreshold
oscillations \cite{Izh01}, or relative refractory effects \cite{Tuc78,LinLon05}
into account. In some of these cases, approximations to the firing statistics in
certain regimes are known and so one can ask whether the additional variables
change the uniqueness  of parameters for a given rate and CV.

Finally, the question of uniqueness of parameters could be also addressed in
models that include a more realistic input beyond the diffusion
approximation. This requires taking into consideration the shot-noise
character of the synaptic input, i.e. the fact that the neuron is driven by
spike trains, as incorporated in Stein's model \citep{Ste65,Ste67}.  The
questions treated here may be worth to be addressed also for neurons which are
subject to colored noise, either caused by a finite synaptic time constant
\citep{BruSer98} or by temporal correlations in the pre-synaptic input
\citep{Lin04}.

\section{Acknowledgements}
We would not have been able to complete parts of our proofs without many
discussions with Jochen Br\"ocker, Gianluigi~Del~Magno, and Tilo~Schwalger;
these are gratefully acknowledged.

{\bf Note added in proof:} The maximum of the CV as a function of the base current for the LIF model has been discussed previously by F. Barbieri and N. Brunel, Front. Comput. Neurosci. {\bf 1}, 5 (2007).

\begin{appendix}
\section{Monotonicity of certain functions of  interest for the LIF} \label{app:ineq}
Here we want to prove that the two functions differences of which occur in
\e{thesisLif3} are monotonically decreasing functions of their arguments. Specifically, we
argue that
\be
e^{a^2} \mbox{erfc}(a)>e^{b^2} \mbox{erfc}(b) \;\;\; \mbox{if} \;\;\; a<b
\label{ineq1}
\ee
and
\be
\int_a^\infty dx \frac{\mbox{erfc}^2(x)}{\mbox{erfc}(a)} > \int_b^\infty dx \frac{\mbox{erfc}^2(x)}{\mbox{erfc}(b)} \;\;\; \mbox{if} \;\;\; a<b
\label{ineq2}
\ee
We start by proving \e{ineq1}. For this purpose, we show that the difference of the left hand
and the right-hand sides is always positive.  Writing
explicitely $\mbox{erfc}(a)=2/\sqrt{\pi}\int_a^{\infty}dx \mbox{ exp}(-x^2)$
and performing the changes of variables $s'=t-a$ and $s"=t-b$, we obtain for
the difference
\ba
&&\frac{2}{\sqrt{\pi}}\left(\int_0^{\infty}ds'e^{-s'^2-2as'}-\int_0^{\infty}ds"e^{-s"^2-2bs"}\right)
=
\frac{2}{\sqrt{\pi}}\int_0^{\infty}ds e^{-s^2}(e^{-2as}-e^{-2bs})>0
\nonumber\\ 
&&\hspace*{10cm} \hfill\mbox{ if } \;\;\; a<b,
\label{mon_decreasing}
\ea
since $s\geq 0$. Therefore,  \e{ineq1} is proven.

The second inequality \e{ineq2}  is equivalent to proving that the function
$$
f(a)=\frac{\int_{a}^{\infty}dx e^{x^2 }\mbox{erfc}^2(x)}{\mbox{erfc}(a)} 
$$
is a monotonically decreasing function of $a$. For this to hold true, the derivative of $f(a)$ should
be negative for all $a$, i.e. 
\be
\frac{df}{da}=\frac{2}{\sqrt{\pi}\mbox{erfc}^2(a)}\int_{a}^{\infty} dx\; \left(e^{x^2 -a^2}\mbox{erfc}^2(x)- e^{a^2 -x^2}\mbox{erfc}^2(a)\right)<0. 
\ee

The prefactor is positive and the integrand is strictly negative for all
$x>a$. The latter can be seen by multiplying the integrand by $e^{x^2 +a^2 }$
from which we obtain $(e^{x^2 }\mbox{erfc}(x))^2-(e^{a^2 }\mbox{erfc}(a))^2$,
which is negative by virtue of \e{ineq1}.  The proof of \e{ineq2} is
therefore completed.

\section{Some properties of the rate in the LIF model} \label{app:rate_props}
Here we show that for the white-noise driven LIF model the derivative of the
rate with respect to the mean input is always positive. Further we calculate
the limits of the rate for the mean input approaching minus and plus infinity.

We first want to prove that
\be
\partial{r}/\partial{\mu}>0.
\label{drdmuLIF}
\ee
Eqs. (\ref{meanISIlif}) and (\ref{a&b}) imply that
\begin{equation}
\partial{\langle T \rangle}/\partial{\mu} = \frac{1}{\sqrt{2D}}\left(
\partial{\langle T \rangle}/\partial{a} + \partial{\langle T
\rangle}/\partial{b} \right).
\end{equation}
Now, 
\be
\label{drdmu}
\partial{\langle T \rangle}/\partial{a} + \partial{\langle T \rangle}/\partial{b}=\sqrt{\pi}(e^{b^2} \mbox{ erfc}(b) - e^{a^2} \mbox{ erfc}(a)).
\ee
We thus obtain:
\be
\frac{\partial \lr{T}}{\partial \mu}=\sqrt{\frac{\pi}{2D}}(e^{b^2} \mbox{ erfc}(b) - e^{a^2} \mbox{ erfc}(a))<0,
\ee
since $a <b$ and by virtue of \e{mon_decreasing}.
\e{drdmuLIF} is therefore proved.

Now let us prove that for the LIF one has 
\be 
\label{limits_of_rate}
\lim_{\mu \rightarrow \infty}r=\lim_{\mu \rightarrow
  \infty}\frac{1}{\lr{T}}=\infty \;\;\; \mbox{and}\;\;\; \lim_{\mu \rightarrow
  -\infty}r=\lim_{\mu \rightarrow - \infty}\frac{1}{\lr{T}}=0.
\label{lim1}
\ee
for which it suffices to show that the mean interval approaches zero or
infinity as $\mu$ goes to plus or minus infinity, respectively. The integrand
in the integral expression for the mean interval \e{meanISIlif} is a
monotonically decreasing function as shown above in \e{mon_decreasing}; with
this property we can estimate
\be
 \sqrt{\pi}(b-a) \mbox{erfc}(b) e^{b^2} \le\lr{T}     \le
 \sqrt{\pi}(b-a) \mbox{erfc}(a) e^{a^2},
\ee
which is equivalent to
\be
\mbox{erfc}(b) e^{b^2} \le  \sqrt{\frac{2D}{\pi}} \frac{\lr{T}}{v_{th}-v_{r}}  \le
 \mbox{erfc}(a) e^{a^2}.
\ee
For $\mu\to\infty$ both $a,b\to\infty$ and the functions on the left and right
hand sides go to zero (see eq.(7.1.23) by \cite{AbrSte70}).  Thus, we obtain in this
limit what proves the first of our limit cases in \e{limits_of_rate}:
\be 
\lim\limits_{\mu\to \infty} \lr{T}=0.  
\ee 
In the opposite limit of $\mu\to -\infty$, both $a,b\to-\infty$; the
complementary error function attains a finite value in this limit
($\lim_{x\to-\infty}$ erfc$(x)=2$) and the exponential functions then yield a
divergence of both sides yielding
\be 
\lim\limits_{\mu\to -\infty} \lr{T}=\infty.  
\ee 
which proves the second  of the asserted limit cases in \e{limits_of_rate}. 
\end{appendix}

\bibliographystyle{apalike}

\end{document}